\documentclass{eas}
\usepackage{graphicx}
\begin{document}

\TitreGlobal{Mass Profiles and Shapes of Cosmological Structures}

%%-----------------------------
%%      the top matter
%%-----------------------------
\title{Mass Accretion Histories \& Density Profiles of $\Lambda$CDM Clusters}

%\footnote{Based on Tasitsiomi, Kravtsov, Gottloeber and Klypin, ApJ 607,125 (2004)}
%
\author{Argyro Tasitsiomi}\address{Department of Astronomy \& Astrophysics, Kavli Institute for Cosmological Physics, University of Chicago, 5640 S. Ellis Ave, Chicago, IL 60637}
\secondaddress{Based on Tasitsiomi, Kravtsov, Gottloeber, and Klypin 2004, ApJ, 607, 125}

%\address{Department of Astronomy \& Astrophysics, Kavli Institute for Cosmological Physics, University of Chicago, 5640 S. Ellis Ave, Chicago, IL 60637}
%
\runningtitle{MAHs \& Profiles of Clusters }
\setcounter{page}{23}
% Keep this line, even if the page will be settled afterwards..
\index{Argyro Tasitsiomi}
% Repeat the authors here, this will help to make the final index
%\maketitle
\begin{abstract} 
  We analyze the mass accretion histories (MAHs) and density profiles
  of cluster-size halos in a flat $\Lambda$CDM cosmology. 
%  Most MAHs consist of 
%  an early, merger-dominated
%  mass increase followed by a more gradual, accretion-dominated
%  growth.  For some clusters the intense merger activity and rapid
%  mass growth continue until the present-day epoch. For these clusters we find that 
%  MAH fits   
  We find that these MAHs are very diverse, and in order to fit all of them 
  we generalize 
  the MAH fit found in previous systematic studies of predominantly galactic halos.
  Moreover, we find that the concentration of the density
  distribution is tightly correlated with the halo's MAH and with its
  formation redshift.  During the early period of fast mass growth the
  concentration remains approximately constant and low $c_{\rm
    v}\approx 3-4$, 
    while during the slow accretion stages the
  concentration increases with decreasing redshift as $c_{\rm
    v}\propto (1+z)^{-1}$.  
    We consider fits of three widely discussed
  analytic density profiles to the simulated clusters.  
  We find that there is no unique best
  fit for all the systems. At the same time, if a
  cluster is best fit by a particular analytic profile at $z=0$, the
  same is usually true at earlier epochs out to $z\sim 1-2$. The local
  logarithmic slope of the density profiles at $3 \%$ of the virial
  radius ranges from $-1.2$ to $-2.0$.
%  , a remarkable diversity for the
%  relatively narrow mass range of our cluster sample.  Interestingly,
   In addition, the logarithmic slope becomes shallower
   with decreasing radius without reaching an asymptotic value down to
   the smallest resolved scale ($< 1\%$ of the virial radius).
%  We do not find a clear correlation of the inner slope with the
%  formation redshift or the shape of the halo's MAH. We do find,
%  however, that 
   During the early MAH period of rapid mass growth the density
  profiles can be well described by a single power law $\rho(r)\propto
  r^{-\gamma}$ with $\gamma\sim 1.5-2$.  The relatively shallow power
  law slopes result in low concentrations at these stages of
  evolution, as the scale radius where the density profiles reaches
  the slope of $-2$ is at large radii.  This indicates that the inner
  power law like density distribution of halos is built up during the
  periods of rapid mass accretion and active merging, while the outer steeper
  profile is formed when the mass accretion slows down. 
%  To check the
%  convergence and robustness of our conclusions, we resimulate one of
%  our clusters using eight times more particles and twice better force
%  resolution.  We find good agreement between the two simulations in
%  all of the results discussed in our study.
\end{abstract}
%\begin{abstract}
%We review briefly results presented in Tasitsiomi, Kravtsov, Gottloeber and Klypin, ApJ 607, 125 (2004)
%\end{abstract}
\maketitle
%
%%-----------------------------
%%      your text
%%-----------------------------
\section{Mass Accretion Histories}
\begin{figure}[h]
   \centering
   \includegraphics[width=9cm]{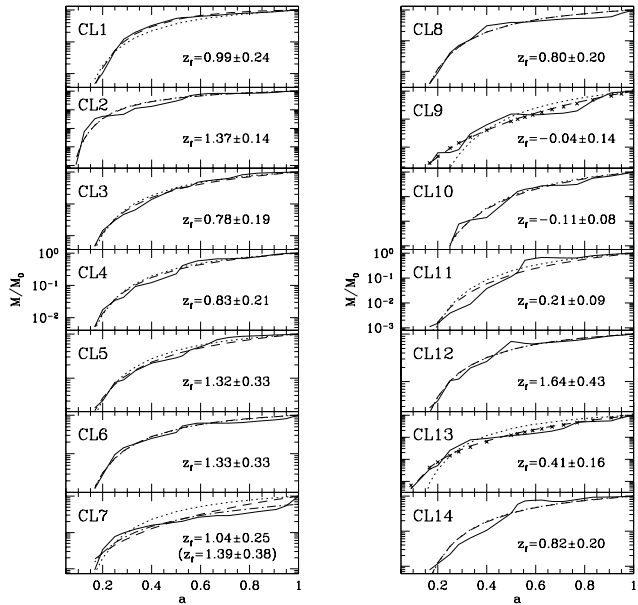}
      \caption{
      Mass accretion histories of  cluster halos (see text for details). 
%  Also shown are the analytic fits of Eq.~(\ref{eq:our_function}),
%  $M(\tilde{a})/M_0= \tilde{a}^p\exp[-\alpha(\tilde{a}-1)]$ (where
%  $\tilde{a}\equiv a/a_0$, and $a=(1+z)^{-1}$). The {\it dashed lines}
%  show fits with both $\alpha$ and $p$ varied, while the {\it dotted
%    lines} show the fits with the parameter $p$ fixed to zero. The
%  formation redshift $z_{f}$, given by Eq.~(\ref{eq:form_z}) in terms
%  of $\alpha$, is shown in the legend of each panel.  The best fit
%  $\alpha$ obtained from the two fits is nearly identical in all
%  cases, except for CL9 and CL13.  For CL9 and 13 we also plot the fit
%  obtained using Eq.~(\ref{eq:vandenbosch}) ({\it crosses}).  Finally,
%  to show the effect of a recent major merger, for CL7 we plot the
%  fit assuming an epoch of observation $a_{0}=0.95$ rather than
%  $a_{0}=1$, used for all the other fits ({\it dash-dotted line}). The
%  $z_f$ value obtained in this case is given in the parentheses.
}
       \label{figure_mafig}
   \end{figure}

      {\bf Figure \ref{figure_mafig}.} The solid line in each panel is the MAH of the corresponding cluster halo. To  construct 
these curves for each one of the 14 halos identified at z=0, we track down its  
most massive progenitor in the various redshifts for which there is a simulation output.The 
dashed curves are the best fits of the formula we propose in Tasitsiomi et al. (2004).
The functional  form proposed  by Wechsler et al. (2002)  based on the 
MAHs of mostly  galactic  halos is shown  with the dotted curves.  Even though  it is a 
good fit for most cluster halos, it  does not seem adequate for all clusters. The clusters which are not
described well by this fit  have MAHs which are close to a power law in the scale factor.
Also given is the formation 
redshift, $z_{f}$, of each halo  calculated via the logarithmic mass accretion rate. Clearly, cluster halos 
are newly formed objects -- some are still forming ($z_{f}<0$). Despite the large MAH diversity, there 
is a typical MAH; it consists of 2 phases, an early ($z>z_{f}$), major merger dominated phase, and a 
later ($z<z_{f}$), slower accretion  phase.

\section{Is there a universal density profile and/or inner slope?}
\begin{figure}[h]
   \centering
   \includegraphics[width=4cm]{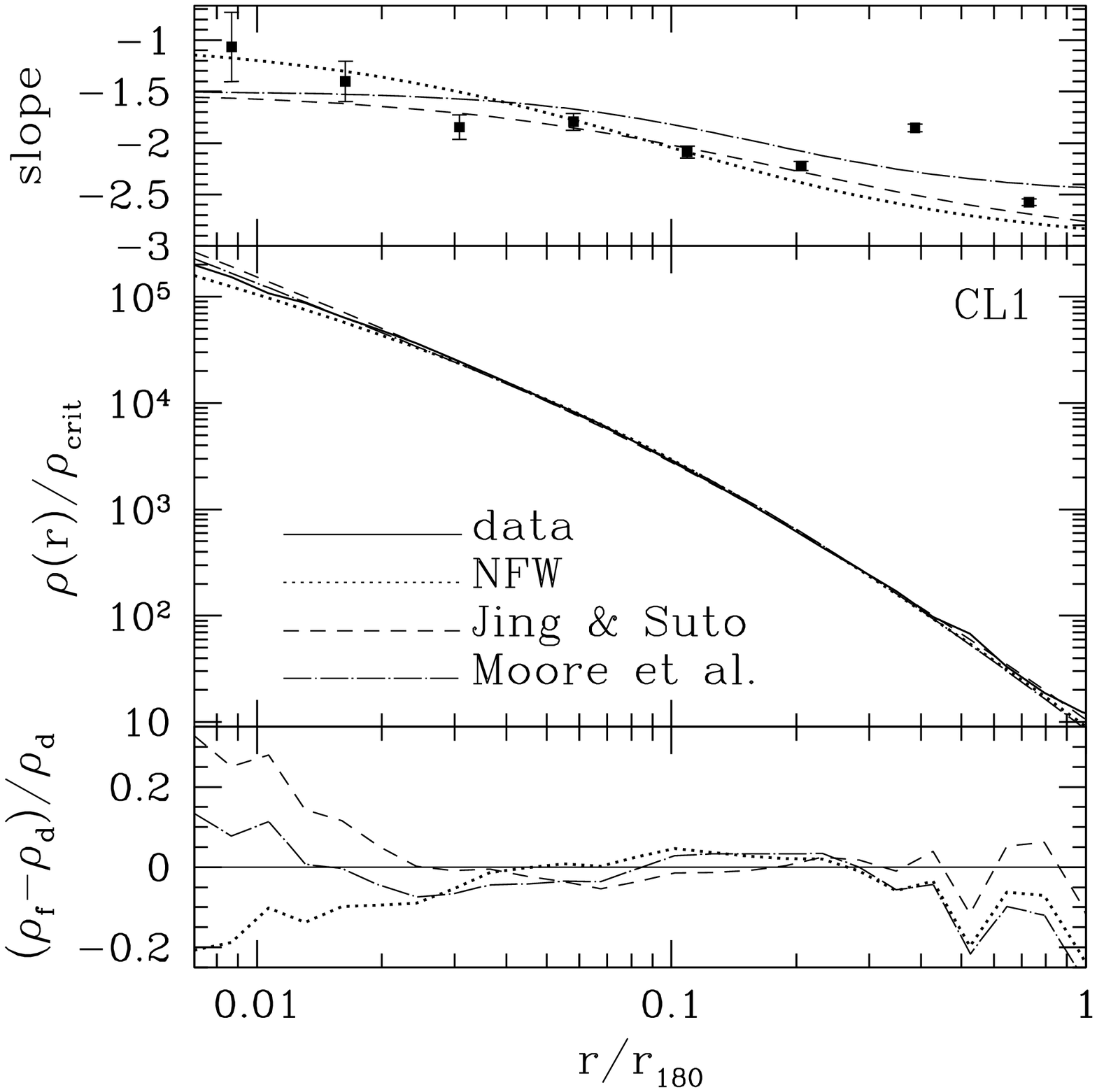}\includegraphics[width=4cm]{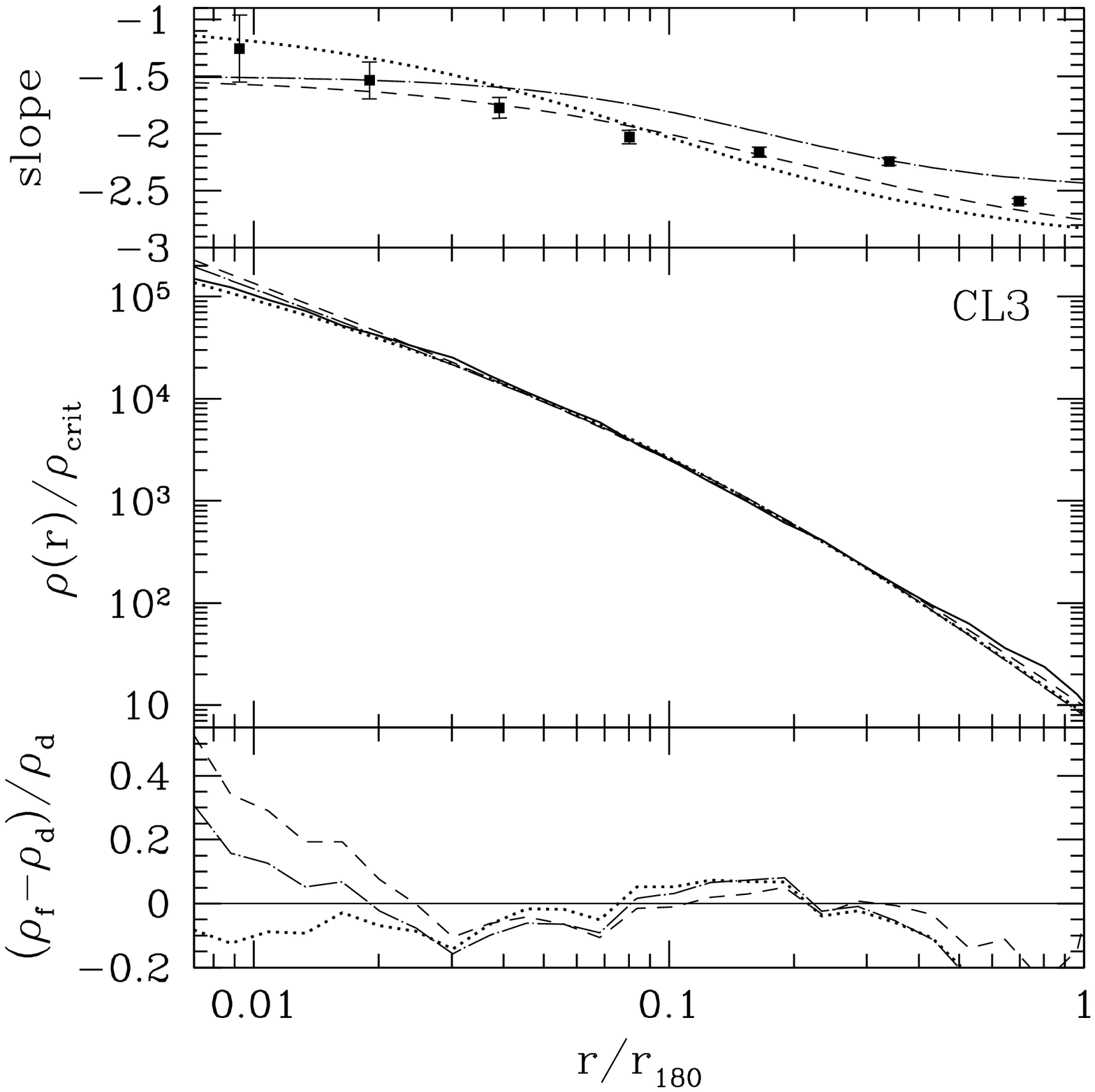}\includegraphics[width=4cm]{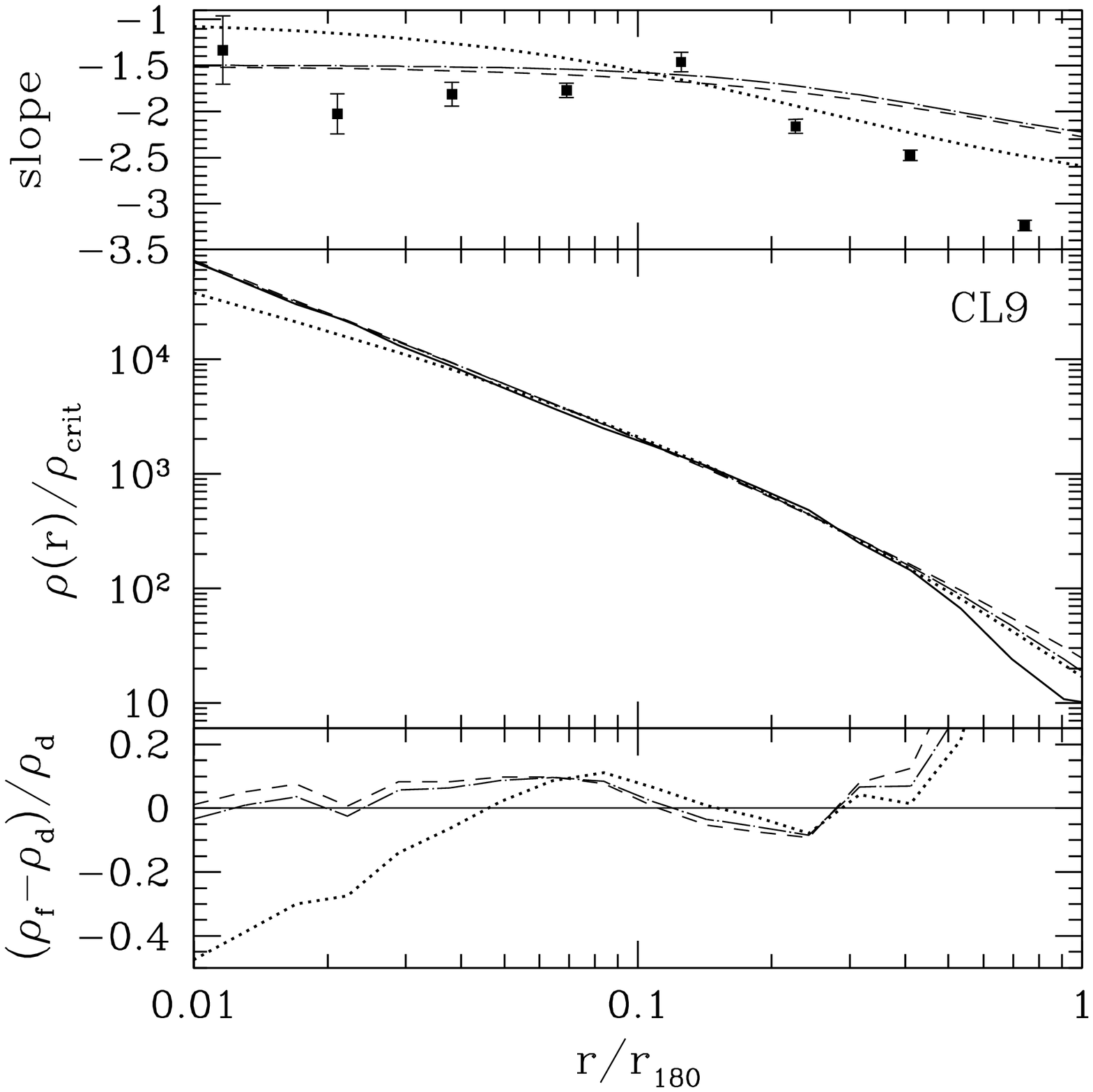}
      \caption{Density profiles, slopes,  and analytical fits for 3 of the clusters.}
\label{figure_profile}
\end{figure}

{\bf Figure \ref{figure_profile}.}  Density profiles and three profile analytical fits for three of the simulated clusters (middle panel of each figure). The 
bottom 
panels show the fractional deviation of the three fits from the data, and the 
top panels show the local logarithmic profile slope as measured from the density profile (points), as well as the predictions 
for the running of the slope from the three analytic fits. We find that there is no unique best fit analytic form for all clusters. CL1 is 
representative of clusters best fit by the Moore et al. (1998)  and CL3 by the Navarro, Frank, and White (NFW; 1996)  profile. In accordance with the large MAH diversity, the 
concentration indices calculated from the best fits vary from 2.3 to 14.7 for a relatively narrow mass range. The logarithmic slope obtained 
by averaging between the smallest resolved radius and 3$\%$ of the virial radius is equally diverse, varying  from -1.23 to -2.01. Opposite to 
previous claims, there are no indications for an asymptotic slope being reached, at least not down to our smallest resolved radius. We do not 
find a correlation between the inner slope and  $z_{f}$, or the shape of the MAH. What we do find is shown at the
right panel: shown is one of the clusters that are currently undergoing rapid mass accretion and, in our definitions, has not formed yet. These 
objects which are still in their first MAH phase, when the inner part of the halo is formed, are single 
power law like over a large range of scales, and have accordingly low concentration indices.  Note that CL9 is one of the objects  not well described 
by the  MAH formula of Wechsler et al. (2002). 
\section{Cluster profile evolution}
\begin{figure}[h]
   \centering
   \includegraphics[width=9cm]{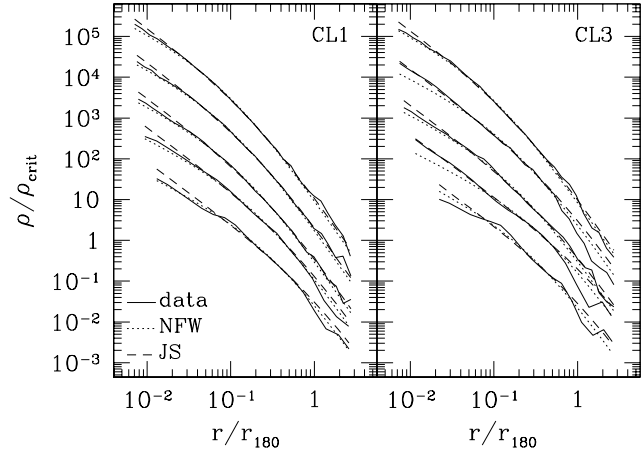}
    \vspace{-1cm}
         \caption{Cluster profile evolution.}
\label{evolution}
\end{figure}

{\bf Figure \ref{evolution}.} Top to bottom, density profile at z=0, 0.2, 0.4, 1, and 1.5 (solid lines) for two different cluster 
halos (left and right panels). The profiles at $z>0$ are scaled down by a factor of 10 with respect 
to each other.  Also shown are the best fit NFW (dotted lines)  and Jing \& Suto (JS; 2000) (dashed 
lines)  profiles.  The inner profile is set up early on,  in most cases during the period of rapid, 
major merger dominated accretion, it is almost a power law (same as CL9 in Figure \ref{figure_profile}),  and 
remains fairly intact when the outer parts of the halo are built during the second, milder mass 
accretion era. In most cases, as those shown in the figure,  the best fit analytic form at z=0, is 
also the best fit in the past. For example, for the cluster shown in the left panel, the best fit 
analytic form is always the JS density profile, whereas for the cluster on the right panel it is the NFW.
\section{Cluster concentrations}
\begin{figure}[h]
   \centering
   \includegraphics[width=7cm]{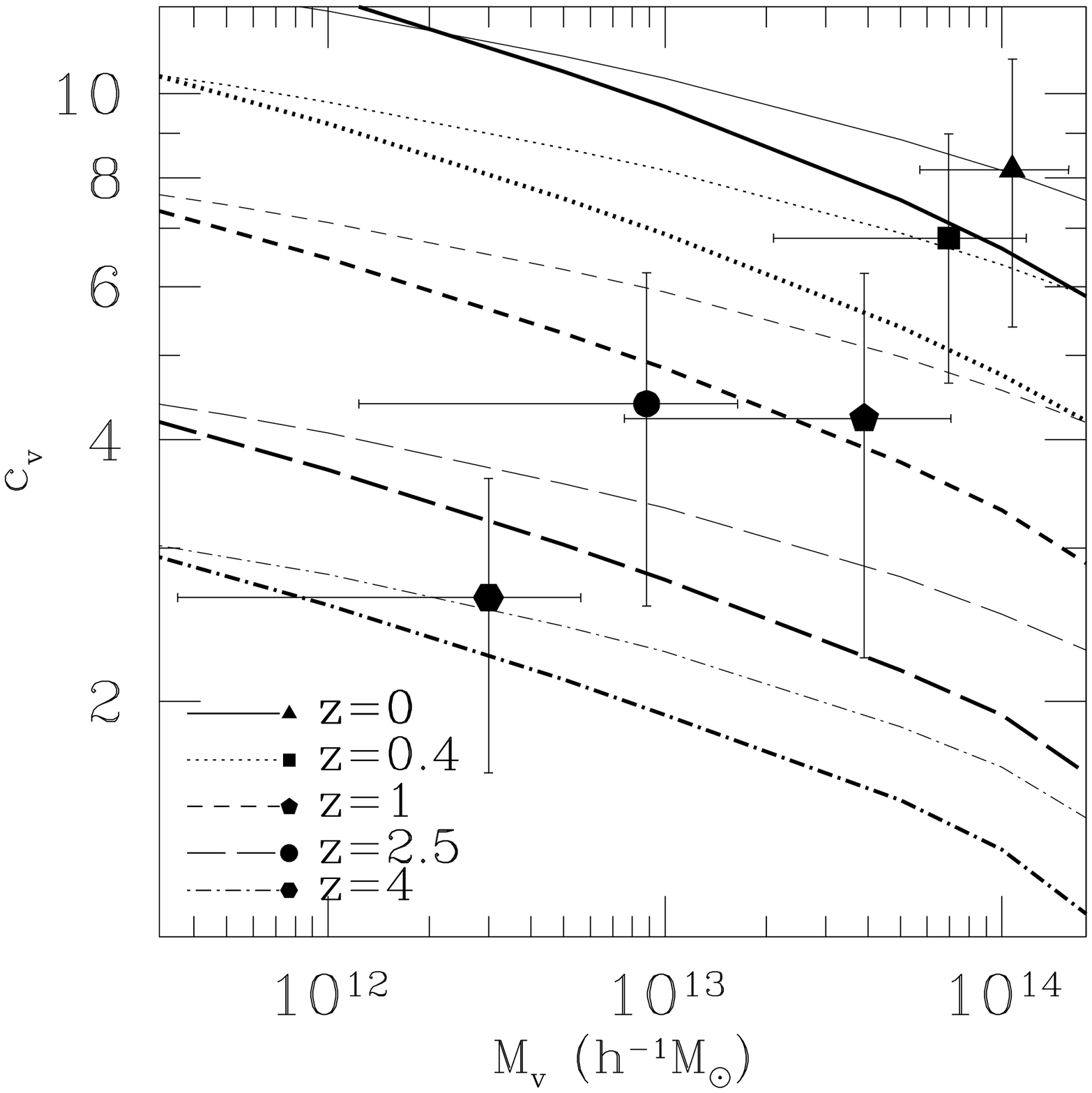}\includegraphics[width=7cm]{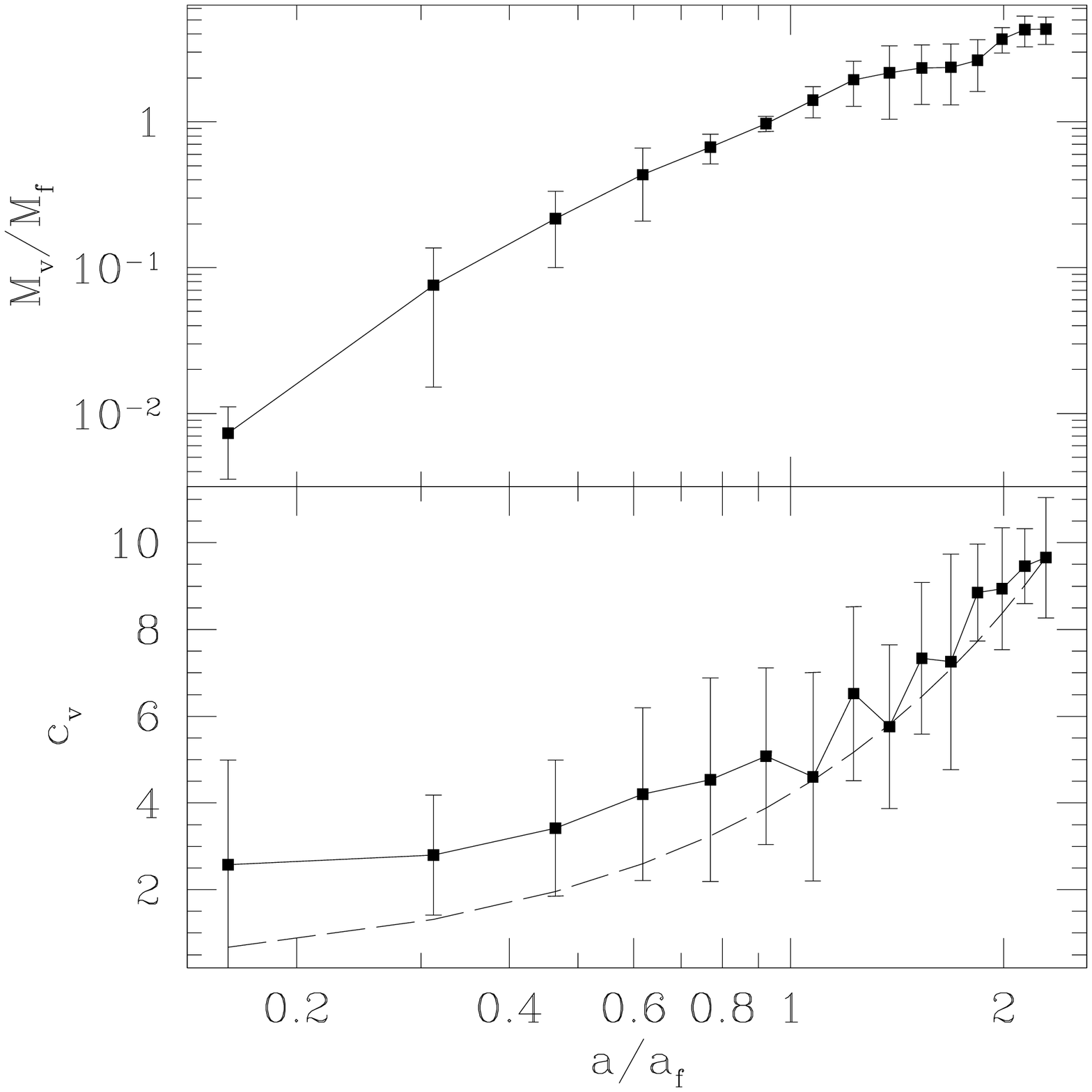}
         \caption
{ Concentration evolution.}
\label{concentrations}
\end{figure}

{\bf Figure \ref{concentrations}}. {\it Left panel:} 
Median concentration vs. virial mass at different redshifts for the progenitors of clusters in 
our sample (points). The vertical error bars represent the 1-$\sigma$  scatter in concentration, while the 
horizontal ones show the mass range of the halos at each epoch. The predictions of  well known  
and mostly based on galactic size halo samples c(M$_{v}$,z) recipes are shown with thick (Bullock et al., 2001)
 and thin (Eke, Navarro, and Steinmetz, 2001) lines.
{\it Right panel:} Average MAH (top panel) and average concentration of cluster progenitors (bottom panel) as a function 
of scale factor in units of the formation scale factor, $a_{f}$. The error bars in both panels show the 1- $\sigma$ spread 
around the mean. The figure shows that the concentration of cluster halos relate tightly to the MAH: during the 
period of fast mass growth the concentration remains approximately constant and low (~3-4), whereas during the 
slow accretion period the concentration increases with decreasing z as $(1+z)^{-1}$. 

%%-----------------------------
%%      your bibliography
%%-----------------------------
%In preparing the reference list please adhere to the following format.
% Attention should be paid to the order of the items in each reference
% and to the punctuation used. Please see Sect. 4 in the User's Guide
% that comes with the new macro package.

%Bohr, N., Einstein, A., & Fermi, E. 1992, MNRAS, 301, 257 (BEF)
% Curie, M., & Curie, P. 1991, A&A, 248, 612
% de Gaulle, C. 1996, Solar Phys. (Oxford: Oxford Univ. Press)
% Heisenberg, W., & West, C. N. 1993, Australian J. Phys., 537, 36  (Paper III)
% Laurel, S., & Hardy, O. 1994, Active Galactic Nuclei, in The Evolution
% and Distribution of Galaxies, ed. W. Churchill, F. D. Roosevelt, & J.
% Stalin (New York: Wiley), 210

\end{document}